\documentclass[a4paper,11pt]{article}

\usepackage{jheppub}

\usepackage{graphicx}
\usepackage{multirow}
\usepackage{bm}
\usepackage{bbm}
\usepackage{color}
\usepackage{slashed}

\title{ Next-to-next-to-leading-order QCD corrections to $J/\psi$ plus $\eta_c$ production at the $B$ factories}

\author[a,b]{Xu-Dong Huang,}
\emailAdd{huangxd@ihep.ac.cn}

\author[a,b]{Bin Gong,}
\emailAdd{twain@ihep.ac.cn}

\author[a,b]{Jian-Xiong Wang}
\emailAdd{jxwang@ihep.ac.cn}

\affiliation[a]{Institute of High Energy Physics, Chinese Academy of Sciences, 19B Yuquan Road, Shijingshan District, Beijing, 100049, P.R. China}
\affiliation[b]{University of Chinese Academy of Sciences, Chinese Academy of Sciences, 19A Yuquan Road, Shijingshan District, Beijing, 100049, P.R. China}

\abstract{In this paper, we calculate the next-to-next-to-leading-order (NNLO) QCD corrections to $e^+e^- \to J/\psi+\eta_c$ at the $B$ factories. After including the NNLO corrections, the cross section of $e^+e^- \to J/\psi+\eta_c$ is enhanced by about $17\%$, and the perturbative series of the prediction shows the convergent behavior. It is also found that the contributions from bottom quark starts at the $\alpha_s^3$-order, which is about $2.4\%$ of the total prediction. The renormalization scale $\mu_R$ dependence of the cross section is reduced at the NNLO level, but the prediction is sensitive to the charm quark mass $m_c$. By considering the uncertainties caused by renormalization scale $\mu_R$, charm quark mass $m_c$ and the NRQCD factorization scale $\mu_\Lambda$, our prediction shows agreement with the B{\footnotesize A}B{\footnotesize AR} and B{\footnotesize ELLE} measurements within errors.}

\begin{document}

\maketitle

\flushbottom

\section{Introduction}

Since the discovery of $J/\psi$ in 1974, the heavy quarkonium production has been a focus of theoretical and experimental researches, which presents an ideal laboratory for the study of the interaction between quarks in two body systems. It plays an important role in the development of quantum chromodynamics (QCD). Perturbative QCD is essential to calculate the theoretical prediction for the large momentum transfer processes. In order to apply it to the quarkonium production, the colour-evaporation model~\cite{Fritzsch:1977ay, Halzen:1977rs}, the color-singlet model~\cite{Chang:1979nn, Berger:1980ni, Matsui:1986dk} and the nonrelativistic QCD (NRQCD) factorization formalism~\cite{Bodwin:1994jh} have been introduced. Among them, the NRQCD factorization formalism provides a rigorous way to calculate the theoretical prediction perturbatively, whose result can be improved by including higher order corrections of the QCD coupling constant $\alpha_s$ and the heavy-quark relative velocity $v$. It is very successful when applied to many quarkonium production processes, especially the unpolarized cross section of the $J/\psi$ hadroproduction~\cite{Brambilla:2010cs, Andronic:2015wma, Lansberg:2019adr, Chen:2021tmf}. However, there are still some discrepancies between the NRQCD predictions and the experimental measurements of the heavy quarkonium production. To test the NRQCD factorization, it is significant to investigate more processes about heavy quarkonium production.

In 2002, the total cross section of $e^+e^- \to J/\psi +\eta_c$ measured by B{\footnotesize ELLE} at $\sqrt{s}=10.58$ GeV~\cite{Belle:2002tfa} is $\sigma[J/\psi+\eta_c] \times B^{\eta_c}[\geq4] = (33^{+7}_{-6}\pm9)\mathrm{~fb}$, where $B^{\eta_c}[\geq n]$ denotes the branching ratio of $\eta_c$ into $n$ or more charged tracks. This measurement was improved as $\sigma[J/\psi+\eta_c] \times B^{\eta_c}[\geq2] = (25.6\pm2.8\pm3.4)\mathrm{~fb}$~\cite{Belle:2004abn} in 2004. Later in the year 2005, another independent measurement was finished by B{\footnotesize A}B{\footnotesize AR}~\cite{BaBar:2005nic}, and the total cross section is $\sigma[J/\psi+\eta_c] \times B^{\eta_c}[\geq2] = (17.6\pm2.8^{+1.5}_{-2.1})\mathrm{~fb}$. Meanwhile, the calculation at NRQCD leading order (LO) of the QCD coupling constant $\alpha_s$ and the charm quark relative velocity $v$, gives a theoretical prediction for the total cross section $2.3 \sim 5.5 \mathrm{~fb}$~\cite{Braaten:2002fi,Liu:2002wq,Hagiwara:2003cw}, which is much smaller than the experimental measurements. A lot of theoretical studies have been performed to explain this large discrepancy. The relativistic corrections have been studied by several groups~\cite{Bodwin:2006ke,He:2007te,Bodwin:2007ga}. Some other attempts have also been suggested to solve this discrepancy, such as the light-cone factorization approach~\cite{Ma:2004qf,Bondar:2004sv,Bodwin:2006dm} or light-cone sum rules~\cite{Braguta:2008tg,Zeng:2021hwt}. The next-to-leading-order (NLO) QCD correction of the process has been regarded as a breakthrough~\cite{Zhang:2005cha,Gong:2007db}, which can greatly enhance the size of cross section and reduce the large discrepancy. The joint NLO QCD and relativistic correction has been investigated in Refs.~\cite{Dong:2012xx,Li:2013otv}. The improved NLO prediction has been given in Ref.~\cite{Sun:2018rgx} by applying the principle of maximum conformality~\cite{Brodsky:2013vpa,Shen:2017pdu,Wu:2019mky,Huang:2021hzr}, which shows excellent agreement with the experimental measurements.

However, the NLO prediction shows very poor convergence, the relative magnitudes of each order is about $1:1$. The NNLO QCD correction is still important to verify its perturbative property. In 2019, the challenging NNLO correction of this process was calculated in Ref.~\cite{Feng:2019zmt}, however the precision of master integrals is not satisfied. In 2022, a powerful algorithm named {\it Auxiliary Mass Flow} has been pioneered by Liu and Ma~\cite{Liu:2017jxz,Liu:2020kpc,Liu:2021wks}, which can be used to compute the Feynman integrals with very high precision. In this paper, we will calculate the NNLO QCD correction to $e^+e^- \to J/\psi + \eta_c$ with the help of the package \texttt{AMFlow}~\cite{Liu:2022chg} and further include the contribution from bottom quark\footnote{Recently, the author of Ref.~\cite{Feng:2019zmt} update their numerical results in the newest version by using the package \texttt{AMFlow}~\cite{Liu:2022chg} too, where the contribution from bottom quark has been also considered. We will compare their numerical results with ours in the following part.}.

The remaining parts of the paper are organized as follows. In Sec.~\ref{I}, we will present useful formulas for the process $e^+e^- \to J/\psi + \eta_c$ and give a brief description on the calculation procedures. In Sec.~\ref{II}, we will show the numerical results and discussions. Sec.~\ref{III} is reserved as a summary.

\section{Calculation technology} \label{I}

\subsection{Cross section}

Under the NRQCD factorization, the cross section for $e^+(k_1)e^-(k_2) \to J/\psi(p_1) + \eta_c(p_2)$ can be written as
\begin{eqnarray}
d\sigma_{e^+e^- \to J/\psi + \eta_c}=d{\hat \sigma}_{e^+e^- \to (c\bar{c})[n_1]+(c\bar{c})[n_2]}\langle {\cal O}^{J/\psi}(n_1)\rangle \langle {\cal O}^{\eta_c}(n_2)\rangle,\label{nrqcdfact}
\end{eqnarray}
where $d{\hat \sigma}$ are short-distance coefficients (SDCs) and $\langle {\cal O}^{J/\psi}(n_1)\rangle$, $\langle {\cal O}^{\eta_c}(n_2)\rangle$ are long-distance matrix elements (LDMEs). In the lowest-order nonrelativistic approximation, only the color-singlet contribution with $n_1=\, ^3S_1^{[1]}$ and $n_2=\, ^1S_0^{[1]}$ need to be considered, which is discovered to be trivial in present process.

Since LDMEs $\langle {\cal O}^{J/\psi}(n_1)\rangle$ and $\langle {\cal O}^{\eta_c}(n_2)\rangle$ include the nonperturbative hadronization effects, we start from the cross section of two on-shell $(c\bar{c})$-pairs with quantum number $^3S_1^{[1]}$ and $^1S_0^{[1]}$,
which has same SDC with $e^+e^-\rightarrow J/\psi +\eta_c$:
\begin{eqnarray}
d \sigma_{e^+e^- \to (c\bar{c})[n_1]+(c\bar{c})[n_2]}=d{\hat \sigma}_{e^+e^- \to (c\bar{c})[n_1]+(c\bar{c})[n_2]}\langle {\cal O}^{(c\bar{c})[n_1]}(n_1)\rangle\langle {\cal O}^{(c\bar{c})[n_2]}(n_2)\rangle. \label{sdcs}
\end{eqnarray}
Here symbols $\langle {\cal O}^{(c\bar{c})[n_1]}(n_1)\rangle$ and $\langle {\cal O}^{(c\bar{c})[n_2]}(n_2)\rangle$ are related to NRQCD bilinear operators as
 \begin{eqnarray}
\langle {\cal O}^{(c\bar{c})[n_1]}(n_1)\rangle&=&|\langle 0| \chi^{\dagger}{\mathbf \epsilon}\cdot{\mathbf \sigma}\psi|c\bar{c}(n_1) \rangle|^2,\\
\langle {\cal O}^{(c\bar{c})[n_2]}(n_2)\rangle&=&|\langle 0| \chi^{\dagger}\psi|c\bar{c}(n_2) \rangle|^2,
\end{eqnarray}
in which $\langle 0| \chi^{\dagger}{\mathbf \epsilon}\cdot{\mathbf \sigma}\psi|c\bar{c}(n_1) \rangle$ and $\langle 0| \chi^{\dagger}\psi|c\bar{c}(n_2) \rangle$ can be calculated in the NRQCD framework~\cite{Bodwin:1994jh,Czarnecki:1997vz,Beneke:1997jm,Czarnecki:2001zc,Kniehl:2006qw,Hoang:2006ty,Chung:2020zqc}. On the other hand, the l.h.s. of Eq.(\ref{sdcs}) can be calculated directly in perturbative QCD. Hence, the SDC $d{\hat \sigma}_{e^+e^- \to (c\bar{c})[n_1]+(c\bar{c})[n_2]}$ can be extracted using Eq.(\ref{sdcs}). In combination with LDMEs $\langle {\cal O}^{J/\psi}(n_1)\rangle$ and $\langle {\cal O}^{\eta_c}(n_2)\rangle$, we can obtain the cross section of $e^+e^- \to J/\psi + \eta_c$ with Eq.(\ref{nrqcdfact}).

Exclusive production of $e^+e^- \to (c\bar{c})[n_1]+(c\bar{c})[n_2]$ at $B$ factories only involves $s$-channel contribution. In the proceeding process, the $e^+ e^-$ first annihilate into a virtual photon, and then it will decay into two final states. By choosing the Feynman gauge, it is convenient to rewrite the differential cross section as~\cite{CTEQ:1993hwr}:
\begin{eqnarray}
d \sigma_{e^+e^- \to (c\bar{c})[^3S_1^{[1]}]+(c\bar{c})[^1S_0^{[1]}]}=\dfrac{1}{4}\dfrac{1}{2s}\dfrac{1}{s^2}L^{\mu\nu}H_{\mu\nu}d\Phi_2. \label{dsigma}
\end{eqnarray}
where $1/4$ comes from the spin average of the initial $e^+e^-$, $1/2s$ is the flux factor, $1/s^2$ comes from photon propagator and $s=(k_1+k_2)^2$ is the squared center-of-mass energy. $L^{\mu\nu}$ and $H_{\mu\nu}$ are the leptonic tensor and hadronic tensor, respectively. $d\Phi_2$ is the differential phase space for the two-body final state. If we focus on the total cross section, $L^{\mu\nu}$ can be equivalently replaced by $4\pi\alpha s(-\frac{8}{3}g^{\mu\nu})$~\cite{Sun:2021tma, Gong:2009ng}. Thus, the complication of calculation can be greatly reduced. The total cross section of the process $e^+e^- \to (c\bar{c})[^3S_1^{[1]}]+(c\bar{c})[^1S_0^{[1]}]$ can be calculated through
\begin{eqnarray}
\sigma_{e^+e^- \to (c\bar{c})[^3S_1^{[1]}]+(c\bar{c})[^1S_0^{[1]}]}=\frac{\alpha}{12 s^{2}}\sqrt{1-\frac{16m_c^2}{s}}H^{\mu}_{~\mu}, \label{sigma}
\end{eqnarray}
where $H^{\mu}_{~\mu}$ corresponds to the decay of a virtual photon into $(c\bar{c})[^3S_1^{[1]}]$ plus $(c\bar{c})[^1S_0^{[1]}]$.

\subsection{Calculation of the perturbative SDC}

In this section, we will give a brief description on the calculation procedures. First, we apply the package \texttt{FeynArts}~\cite{Hahn:2000kx} to generate corresponding Feynman diagrams and amplitudes for $\gamma^*\to (c\bar{c})[^3S_1^{[1]}]+(c\bar{c})[^1S_0^{[1]}]$ at NNLO in $\alpha_s$. Second, we implement the package \texttt{FeynCalc}~\cite{Mertig:1990an,Shtabovenko:2016sxi} to handle the Lorentz index contraction and Dirac/$SU(N_c)$ traces. Third, we use the Mathematica code developed by Yan-Qing Ma to decompose the Feynman amplitudes into different Feynman integral families. Last, we employ the package \texttt{AMFlow}~\cite{Liu:2022chg} to calculate the Feynman integral families with the help of \texttt{Kira}~\cite{Klappert:2020nbg}, where the package \texttt{Kira} is used to do the integration-by-parts (IBP) reduction.

\begin{figure}[htbp]
\centering
\includegraphics[width=0.8\textwidth]{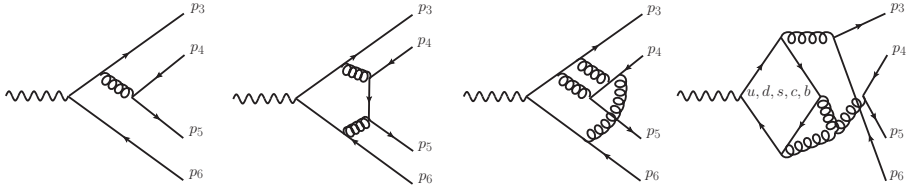}
\caption{Some representative Feynman diagrams for $\gamma^*\to (c\bar{c})[^3S_1^{[1]}]+(c\bar{c})[^1S_0^{[1]}]$.} \label{feynnlo}
\end{figure}

During the calculation, there are nearly 2000 two-loop diagrams for the $\gamma^*\to (c\bar{c})[^3S_1^{[1]}]+(c\bar{c})[^1S_0^{[1]}]$. Some representative Feynman diagrams up to two-loop order are plotted in Figure~\ref{feynnlo}, where $p_3=\frac{p_1+q_1}{2}$, $p_4=\frac{p_1-q_1}{2}$, $p_5=\frac{p_2+q_2}{2}$, $p_6=\frac{p_2-q_2}{2}$ and $p_1^2=p_2^2=(2m_c)^2$. The momenta $q_1$ and $q_2$ denote the relative momenta between the quark and antiquark in the $(c\bar{c})$-pairs. The last Feynman diagram is also called "light-by-light" Feynman diagram, which denotes a special topology. Such topology shows that a closed quark loop is linked with the virtual photon and three gluons. The contributions of "light-by-light" Feynman diagram from the light quark loops can be ignored, since the sum of electric charges of all the light quarks is zero, i.e., $e_u+e_d+e_s=0$. The total NNLO amplitudes can be decomposed into roughly 150 Feynman integral families. These Feynman integral families contain about 10000 Feynman integrals. Due to the powerful package \texttt{AMFlow}, we can compute these Feynman integrals with very high precision. We demand 20-digit precision for each Feynman integral family, and the final numerical result achieve at least 10-digit precision.

In order to obtain the final NNLO correction, we adopt the conventional dimensional regularization approach with $d=4-2\epsilon$ to regulate ultraviolet (UV) and infrared (IR) divergences. Feynman diagrams with a virtual gluon line connected with the quark pair in a meson contain Coulomb singularities, which are shown as power divergence in the IR limit of the relative momentum and can be taken into the $c\bar{c}$ wave function renormalization~\cite{Kramer:1995nb,Gong:2007db}. In our calculation, we set the relative momenta ($q_1$ and $q_2$) between the quark and antiquark in the $(c\bar{c})$-pairs to zero before performing loop integration. The Coulomb divergence vanishes in the calculation under dimensional regularization. The UV divergences are removed through renormalization. We renormalize the heavy quark field and the heavy quark mass in the on-shell (OS) scheme. The coupling constant $\alpha_s$ is renormalized in the $\overline{\rm MS}$ scheme. More explicitly, the amplitudes are renormalized according to
\begin{eqnarray}
&&\mathcal{A}(\alpha_s,m_Q)=\nonumber \\
&&Z_{2,c}^2\bigg[\mathcal{A}_{bare}^{0l}(\alpha_{s,bare},m_{Q,bare})+\mathcal{A}_{bare}^{1l}(\alpha_{s,bare},m_{Q,bare})+\mathcal{A}_{bare}^{2l}(\alpha_{s,bare},m_{Q,bare})\bigg], \label{rn2l1}
\end{eqnarray}
where the $\mathcal{A}^{il}_{bare}|_{i=0,1,2}$ are the tree, one-loop and two-loop bare amplitudes, respectively. $Z_{2,c}$ is the on-shell wave-function renormalization constant for charm quark. The bare mass is renormalized as $m_{Q,bare} = Z_{m,Q} m_Q$, where $Z_{m,Q}$ is the on-shell mass renormalization constants for heavy-quarks. The bare coupling constant is renormalized as
\begin{eqnarray}
\alpha_{s,bare}=\left(\frac{e^{\gamma_E}}{4\pi}\right)^{\epsilon}\mu_R^{2\epsilon}Z_{\alpha_s}^{\overline{\rm MS}}\alpha_s(\mu_R), \label{alphasbare}
\end{eqnarray}
which corresponds to the $\overline{\rm MS}$ scheme with $n_f$ active favors. Here $\mu_R$ is the renormalization scale and $Z_{\alpha_s}^{\overline{\rm MS}}$ is the $\overline{\rm MS}$ coupling constant renormalization constant. The two loop renormalization constants are computed in Refs.~\cite{Broadhurst:1991fy,Bekavac:2007tk,Czakon:2007ej,Czakon:2007wk,Fael:2020bgs}. The renormalized $\mathcal{A}(\alpha_s,m)$ can be obtained by expanding the r.h.s. of Eq.~(\ref{rn2l1}) over renormalized quantities to $\mathcal{O}(\alpha_s^4)$, i.e.,
\begin{eqnarray}
\mathcal{A}(\alpha_s,m_Q)&=&\mathcal{A}^{0l}(\alpha_s,m_Q)+\left(\frac{e^{\gamma_E}}{4\pi}\right)^{-\epsilon}\mathcal{A}^{1l}(\alpha_s,m_Q)+\left(\frac{e^{\gamma_E}}{4\pi}\right)^{-2\epsilon}\mathcal{A}^{2l}(\alpha_s,m_Q)+\mathcal{O}(\alpha_s^4). \nonumber \\
\end{eqnarray}
where the $\mathcal{A}^{il}|_{i=0,1,2}$ are the tree, one-loop and two-loop renormalized amplitudes, respectively. It should be noted that the prefactor $[e^{\gamma_E}/(4\pi)]^{-n\epsilon}$ have been introduced in order to avoid unnecessary $\gamma_E-\ln(4\pi)$ terms. The loop integrals are computed with the measure $\mu_R^{2\epsilon}d^dk/(2\pi)^d$, and the corresponding renormalization constants ($Z_{2,c}$, $Z_{m,Q}$ and $Z_{\alpha_s}^{\overline{\rm MS}}$) can be found in Refs.~\cite{Barnreuther:2013qvf,Tao:2022qxa}. Thus, the total cross section can be written as
\begin{eqnarray}
&&\sigma_{e^+e^- \to (c\bar{c})[^3S_1^{[1]}]+(c\bar{c})[^1S_0^{[1]}]}\nonumber \\
&=&\kappa\bigg|\mathcal{A}^{0l}(\alpha_s,m_Q)+\left(\frac{e^{\gamma_E}}{4\pi}\right)^{-\epsilon}\mathcal{A}^{1l}(\alpha_s,m_Q)+\left(\frac{e^{\gamma_E}}{4\pi}\right)^{-2\epsilon}\mathcal{A}^{2l}(\alpha_s,m_Q)+\mathcal{O}(\alpha_s^4)\bigg|^2 \nonumber \\
&=& \kappa(f_0+f_1+f_2)+\mathcal{O}(\alpha_s^5), \label{sigmaA}
\end{eqnarray}
where $\kappa=\frac{\alpha}{12 s^{2}}\sqrt{1-\frac{16m_c^2}{s}}$. The $f_i(i=0,1,2,)$ denote the squared amplitudes corresponding to $\alpha_s^2$-, $\alpha_s^3$- and $\alpha_s^4$-orders respectively, which are
\begin{eqnarray}
f_0&=&\bigg|\mathcal{A}^{0l}(\alpha_s,m_Q)\bigg|^2,\\
f_1&=&\left(\frac{e^{\gamma_E}}{4\pi}\right)^{-\epsilon}2{\rm Re}\bigg[\mathcal{A}^{1l}(\alpha_s,m_Q)\mathcal{A}^{0l,*}(\alpha_s,m_Q)\bigg],\\
f_2&=&\left(\frac{e^{\gamma_E}}{4\pi}\right)^{-2\epsilon}\bigg\{2{\rm Re}\bigg[\mathcal{A}^{2l}(\alpha_s,m_Q)\mathcal{A}^{0l,*}(\alpha_s,m_Q)\bigg]+\bigg|\mathcal{A}^{1l}(\alpha_s,m_Q)\bigg|^2\bigg\}.
\end{eqnarray}
However, there still remains IR divergence in $f_2$, which can be canceled by including the two-loop corrections to $\langle {\cal O}^{(c\bar{c})[^3S_1^{[1]}]}(^3S_1^{[1]})\rangle$ and $\langle {\cal O}^{(c\bar{c})[^1S_0^{[1]}]}(^1S_0^{[1]})\rangle$ in $\overline{\rm MS}$ scheme. At the lowest order in velocity expansion, it can be written as
\begin{eqnarray}
\langle {\cal O}^{(c\bar{c})[^3S_1^{[1]}]}(^3S_1^{[1]})\rangle|_{\overline{\rm MS}}&=&2N_c\bigg[1-\alpha_s^2(\mu_R)\left(\frac{\mu_{\Lambda}^2e^{\gamma_E}}{\mu_R^2 4\pi}\right)^{-2\epsilon}\left(\frac{C_F^2}{3}+\frac{C_F C_A}{2}\right)\frac{1}{2\epsilon}\bigg], \\
\langle {\cal O}^{(c\bar{c})[^1S_0^{[1]}]}(^1S_0^{[1]})\rangle|_{\overline{\rm MS}}&=&2N_c\bigg[1-\alpha_s^2(\mu_R)\left(\frac{\mu_{\Lambda}^2e^{\gamma_E}}{\mu_R^2 4\pi}\right)^{-2\epsilon}\left(C_F^2+\frac{C_F C_A}{2}\right)\frac{1}{2\epsilon}\bigg],
\end{eqnarray}
which can be obtained from Refs.~\cite{Bodwin:1994jh,Czarnecki:1997vz,Beneke:1997jm,Czarnecki:2001zc,Kniehl:2006qw,Hoang:2006ty,Chung:2020zqc}. The factor $\left(\mu_{\Lambda}^2/\mu_R^2\right)^{-2\epsilon}$ is derived by running the scale of $\alpha_s$ from factorization scale $\mu_{\Lambda}$ to renormalization scale $\mu_R$, since the initial results are calculated at the scale $\mu_{\Lambda}$. The factor $[e^{\gamma_E}/(4\pi)]^{-2\epsilon}$ comes from the definition of $\alpha_s$ in the $\overline{\rm MS}$ scheme given in Eq.~(\ref{alphasbare}). Therefore, the SDC can be determined as
\begin{eqnarray}
{\hat \sigma}_{e^+e^- \to (c\bar{c})[^3S_1^{[1]}]+(c\bar{c})[^1S_0^{[1]}]}&=&\frac{\sigma_{e^+e^- \to (c\bar{c})[^3S_1^{[1]}]+(c\bar{c})[^1S_0^{[1]}]}}{\langle {\cal O}^{(c\bar{c})[^3S_1^{[1]}]}(^3S_1^{[1]})\rangle|_{\overline{\rm MS}}\langle {\cal O}^{(c\bar{c})[^1S_0^{[1]}]}(^1S_0^{[1]})\rangle|_{\overline{\rm MS}}} \nonumber \\
&=& \frac{\kappa}{(2N_c)^2}(f_0+f_1+\tilde{f}_2), \label{sigmaB}
\end{eqnarray}
where $\tilde{f}_2$ is
\begin{eqnarray}
\tilde{f}_2=f_2+f_0\alpha_s^2(\mu_R)\left(\frac{\mu_{\Lambda}^2e^{\gamma_E}}{\mu_R^2 4\pi}\right)^{-2\epsilon}\left(\frac{4C_F^2}{3}+C_F C_A\right)\frac{1}{2\epsilon}.
\end{eqnarray}
The resultant $\tilde{f}_2$ is finite, which renders the SDC without any divergences. And it develops an explicit logarithmic dependence on NRQCD factorization scale $\mu_{\Lambda}$ $\sim\ln\frac{\mu_\Lambda^2}{m^2}$~\cite{Feng:2019zmt} at the NNLO level simultaneously.

\section{Phenomenological results} \label{II}

To do the numerical calculation, the input parameters are taken as follows:
\begin{eqnarray}
&& m_b=4.78 {\rm GeV}, \alpha_s(m_{_Z})=0.1179, \sqrt{s}=10.58 {\rm GeV}, \alpha(\sqrt{s})=1/130.9, \\
&& \langle {\cal O}^{J/\psi}(^3S_1^{[1]})\rangle=0.440^{+0.067}_{-0.056} {\rm GeV}^{3},
\langle {\cal O}^{\eta_c}(^1S_0^{[1]})\rangle=0.437^{+0.111}_{-0.104} {\rm GeV}^{3},
\end{eqnarray}
where the bottom pole mass and running QCD coupling constant at scale $m_{_Z}$ are taken from Particle Data Group~\cite{ParticleDataGroup:2022pth}. The QED coupling constant and NRQCD LDMEs are taken from Ref.~\cite{Bodwin:2007ga}. For simplicity, we take the central value of the LDMEs to perform the phenomenological discussion. We use the package \texttt{RunDec3}~\cite{Herren:2017osy} to evaluate the running QCD coupling constant $\alpha_s(\mu_R)$ at three-loop accuracy.

The numerical results of the NNLO QCD corrections to $J/\psi+\eta_c$ production at the $B$ factories with three typical $m_c$ values are
\begin{eqnarray}
\sigma|_{m_c=1.3 {\rm GeV}}&&=136.651\alpha_s^2(\mu_R)+\bigg[(210.237-14.4991n_l)\ln\frac{\mu_R^2}{m_c^2}+16.5325n_l \nonumber \\
&&+297.262\bigg]\alpha_s^3(\mu_R)+\bigg[(242.587-33.4603n_l+1.1538n_l^2)\left(\ln\frac{\mu_R^2}{m_c^2}\right)^2 \nonumber \\
&&+(818.692-31.0801n_l-2.63123n_l^2)\ln\frac{\mu_R^2}{m_c^2}-870.518\ln\frac{\mu_\Lambda^2}{m_c^2}\nonumber \\
&&+124.921-87.7298n_l-2.29574n_l^2\bigg]\alpha_s^4(\mu_R), \label{sigma13}
\end{eqnarray}
\begin{eqnarray}
\sigma|_{m_c=1.5 {\rm GeV}}&&=115.599\alpha_s^2(\mu_R)+\bigg[(177.849-12.2654n_l)\ln\frac{\mu_R^2}{m_c^2}+10.4752n_l \nonumber \\
&&+215.393\bigg]\alpha_s^3(\mu_R)+\bigg[(205.215-28.3055n_l+0.976053n_l^2)\left(\ln\frac{\mu_R^2}{m_c^2}\right)^2 \nonumber \\
&&+(609.319-28.6518n_l-1.66718n_l^2)\ln\frac{\mu_R^2}{m_c^2}-736.409\ln\frac{\mu_\Lambda^2}{m_c^2}\nonumber \\
&&-109.15-74.7989n_l-2.49917n_l^2\bigg]\alpha_s^4(\mu_R), \label{sigma15}
\end{eqnarray}
\begin{eqnarray}
\sigma|_{m_c=1.7 {\rm GeV}}&&=93.0233\alpha_s^2(\mu_R)+\bigg[(143.116-9.87007n_l)\ln\frac{\mu_R^2}{m_c^2}+5.9587n_l \nonumber \\
&&+150.238\bigg]\alpha_s^3(\mu_R)+\bigg[(165.138-22.7776n_l+0.785436n_l^2)\left(\ln\frac{\mu_R^2}{m_c^2}\right)^2 \nonumber \\
&&+(437.037-25.0833n_l-0.948356n_l^2)\ln\frac{\mu_R^2}{m_c^2}-592.593\ln\frac{\mu_\Lambda^2}{m_c^2}\nonumber \\
&&-231.418-59.3073n_l-2.29771n_l^2\bigg]\alpha_s^4(\mu_R), \label{sigma17}
\end{eqnarray}
where $n_l$ is the number of light flavors ($u$, $d$ and $s$). Note that we do not distinguish the contribution from the light-by-light Feynman diagrams in Eqs.~(\ref{sigma13}-\ref{sigma17}). To guarantee our result reliable, we have also reproduced the NNLO corrections to $e^+e^- \to \eta_c + \gamma $ process~\cite{Yu:2020tri}.

\begin{table}[h]
\centering
\begin{tabular}{c c c c c c}
\hline
 &  & $\alpha_s^2$-terms & $\alpha_s^3$-terms & $\alpha_s^4$-terms & ${\rm Total}$ \\
\hline
 \multirow{2}{*}{$\mu_\Lambda=m_c$}  & $\mu_R=2m_c$  & $7.40$ & $7.04+0.13$ & $2.17+0.28$  & $16.61+0.41$  \\
  &$\mu_R=\sqrt{s}/2$  & $5.06$ & $5.57-0.05$ & $3.43-0.08$  & $14.06-0.13$  \\
\hline
 \multirow{2}{*}{$\mu_\Lambda=1 {\rm GeV}$ }  &$\mu_R=2m_c$  & $7.40$ & $7.04+0.13$ & $4.62+0.27$  & $19.06+0.40$  \\
  &$\mu_R=\sqrt{s}/2$  & $5.06$ & $5.57-0.05$ & $4.58-0.09$  & $15.21-0.14$ \\
\hline
\end{tabular}
\caption{The NNLO cross section (in fb) of $e^+e^- \to J/\psi + \eta_c$ with two typical renormalization scales $\mu_R$ under two factorization scale $\mu_\Lambda$ choices. The second numbers appearing in the $\alpha_s^3$-terms, $\alpha_s^4$-terms and ${\rm Total}$ are the contributions from bottom quark.}  \label{muLambdamc}
\end{table}

Assuming $m_c=1.5$ GeV and the factorization scale $\mu_\Lambda=m_c$ (or $\mu_\Lambda=1$ GeV), we show the total cross section for two typical choices of renormalization scale in Table~\ref{muLambdamc}. It shows that the contribution from bottom quark is about $0.9\thicksim2.4\%$ of the total prediction. It is also found that the $\alpha_s^4$-terms of present process is sizable and smaller than $\alpha_s^3$-terms. The perturbative expansion of the NNLO prediction has exhibited a convergent signature. More explicitly, the relative magnitudes of the $\alpha_s^2$-terms: $\alpha_s^3$-terms: $\alpha_s^4$-terms is about $1: 97\%: 33\%$ for the case of $\mu_R=2m_c$ and $1: 109\%: 66\%$ for the case of $\mu_R=\sqrt{s}/2$. As regards to the renormalization scale $\mu_R$ dependence, it is found that $\sigma \in [13.93, 17.02]$ fb for $\mu_R\in[2m_c, \sqrt{s}/2]$ with $\mu_\Lambda=m_c$; i.e., the scale uncertainty of NLO (NNLO) prediction is $\thicksim27\%$ ($18\%$), respectively\footnote{We attempt to eliminate the renormalization scale ambiguity by using the principle of maximum conformality~\cite{Brodsky:2013vpa,Shen:2017pdu,Wu:2019mky,Huang:2021hzr}. Due to the fact that the effective scale $Q_*$ is found to be $0.37$ GeV, which is already in the non-perturbative region. Thus, we have to give up the discussion in present paper.}. Nevertheless, the renormalization scale dependence of the total cross section is improved $\thicksim9\%$ by including NNLO correction.

To study the factorization scale uncertainty, we also present the numerical NNLO prediction at the factorization scale $\mu_\Lambda=1$ GeV in Table~\ref{muLambdamc}. It shows that the $\alpha_s^4$-terms will change about $34\thicksim99.5\%$ comparing with that of the case $\mu_\Lambda=m_c$. It also indicates that our results are consistent with Table I of Ref.~\cite{Feng:2019zmt}. The net small difference are caused by the different choices of bottom quark pole mass and the significant
digits of $\alpha_s(\sqrt{s}/2)$.

\begin{table}[h]
\centering
\begin{tabular}{c c c c c c}
\hline
 &  & $\alpha_s^2$-terms & $\alpha_s^3$-terms & $\alpha_s^4$-terms & ${\rm Total}$($\mu_\Lambda=m_c$) \\
\hline
 \multirow{2}{*}{$m_c=1.3 {\rm GeV}$}   & $\mu_R=2m_c$    & 9.80 & 11.10 & 5.70  & 26.60  \\
 &  $\mu_R=\sqrt{s}/2$  & 5.98 & 7.46 & 5.77  & 19.21 \\
\hline
 \multirow{2}{*}{$m_c=1.5 {\rm GeV}$}   & $\mu_R=2m_c$    & 7.40 & 7.17 & 2.45  & 17.02  \\
 &  $\mu_R=\sqrt{s}/2$  & 5.06 & 5.52 & 3.35  & 13.93 \\
\hline
 \multirow{2}{*}{$m_c=1.7 {\rm GeV}$}   & $\mu_R=2m_c$    & 5.42 & 4.58 & 0.88  & 10.88  \\
 &  $\mu_R=\sqrt{s}/2$  & 4.07 & 3.90 & 1.74  & 9.71 \\
\hline
\end{tabular}
\caption{The NNLO cross section (in fb) of $e^+e^- \to J/\psi + \eta_c$ under three choices of charm quark pole mass. The two upper rows, the two middle rows and the two lower rows correspond to $m_c=1.3$ GeV, $m_c=1.5$ GeV and $m_c=1.7$ GeV, respectively. The $\mu_R=2m_c$ and $\mu_R=\sqrt{s}/2$ are considered. $\mu_\Lambda=m_c$.} \label{diffmc}
\end{table}

In Table~\ref{diffmc}, we list the numerical results of the NNLO prediction with $m_c=1.3$ GeV, $m_c=1.5$ GeV and $m_c=1.7$ GeV, respectively. It indicates the predicted cross section is rather sensitive to the charm mass. By taking $m_c=1.5$ GeV as center value, the relative uncertainties of $\alpha_s^2$-, $\alpha_s^3$- and $\alpha_s^4$-terms are about $59\%$, $91\%$, and $197\%$ for $\mu_R=2m_c$ and $38\%$, $64\%$, and $120\%$ for $\mu_R=\sqrt{s}/2$, respectively. It can be found that the prediction with $m_c=1.5$ GeV is much closer to the B{\footnotesize A}B{\footnotesize AR} measurements and the prediction with $m_c=1.3$ GeV is much closer to the B{\footnotesize ELLE} measurements.

\begin{figure}[htbp]
\includegraphics[width=0.5\textwidth]{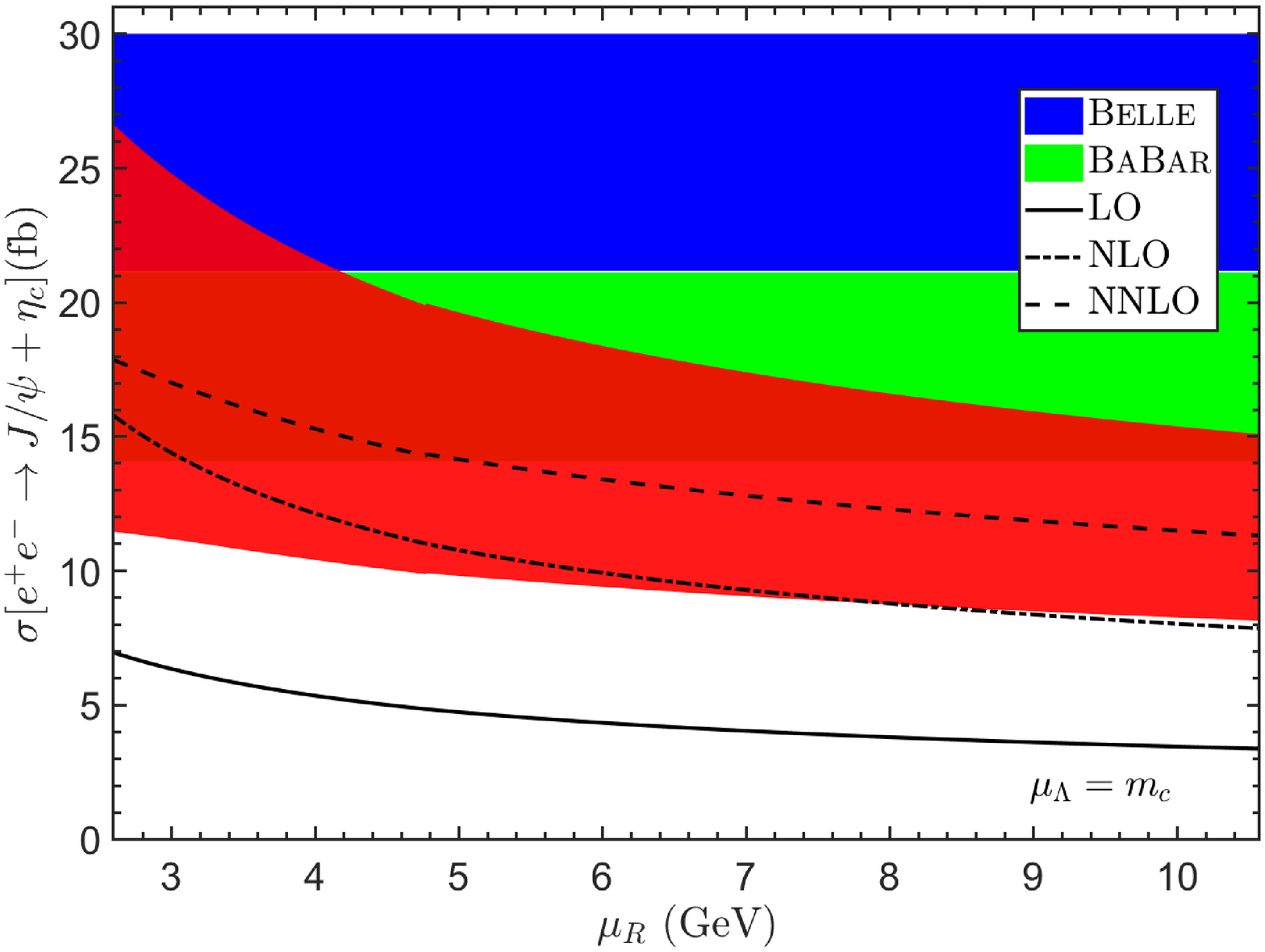}
\includegraphics[width=0.5\textwidth]{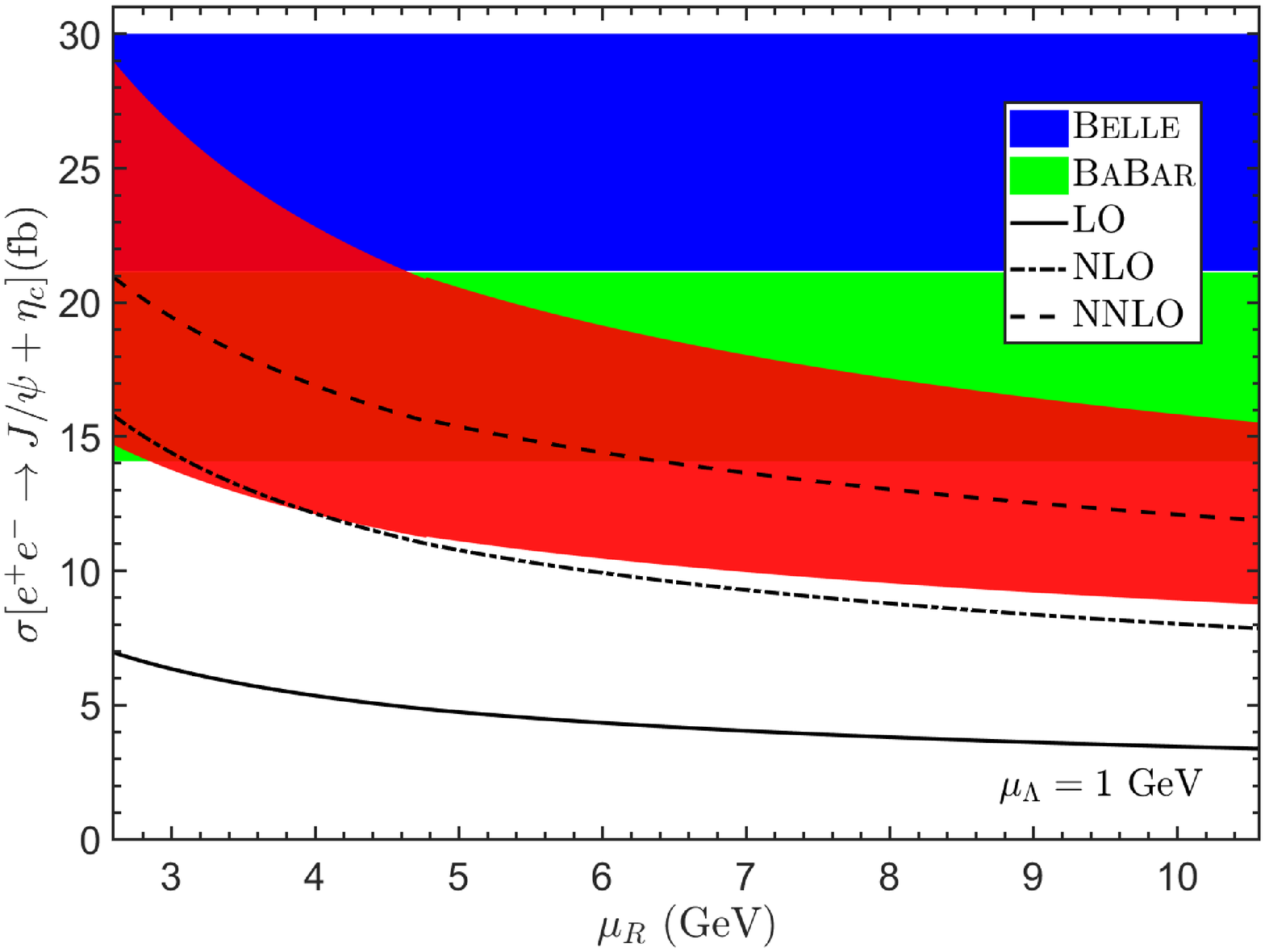}
\caption{The $\mu_R$ dependence of the predicted cross sections at LO, NLO and NNLO levels, respectively. We take $m_c=1.5$ GeV. The red bands represents the uncertainty obtained by varying $m_c \in [1.3, 1.7]$ GeV, where the lower bound corresponds to $m_c=1.7$ GeV and upper bound $m_c=1.3$ GeV. The left figure corresponds to $\mu_{\Lambda}=m_c$, and the right figure corresponds to $\mu_{\Lambda}=1$ GeV.} \label{result}
\end{figure}

In Fig.~\ref{result}, we plot the $\mu_R$ dependence of the predicted cross sections at LO, NLO and NNLO levels, respectively. The three black lines are obtained by taking $m_c=1.5$ GeV. The red bound denotes the uncertainty from $m_c$ within $[1.3, 1.7]$ GeV. The factorization scale are taken as $\mu_{\Lambda}=m_c$ and $\mu_{\Lambda}=1$ GeV, respectively. Fig.~\ref{result} shows that: 1) the NNLO prediction has a milder dependence on the renormalization scale than the NLO prediction in the case of $\mu_\Lambda=m_c$; 2) the NNLO prediction with $\mu_{\Lambda}=1$ GeV are much closer to the experimental value, but much larger $\mu_R$ dependence than that in the case of $\mu_\Lambda=m_c$; 3) the predicted results with $\mu_R=2m_c$ agree with the experimental results better than the results with $\mu_R=\sqrt{s}/2$. Comparing our Fig.~\ref{result} with Fig.~2 of Ref.~\cite{Feng:2019zmt}, it can be found that our result is smaller than their result in the case of $\mu_\Lambda=m_c$~\footnote{ It can be found from Eqs.~(3-11) of Ref.~\cite{Feng:2019zmt} that the numerical result of $\sigma_{\rm NNLO,\mu_\Lambda=1 GeV}$ is larger than that of $\sigma_{\rm NNLO,\mu_\Lambda=m_c}$. Fig.~\ref{result} of our work shows the same conclusion. However, Fig.~2 of Ref.~\cite{Feng:2019zmt} shows that the numerical result of $\sigma_{\rm NNLO,\mu_\Lambda=m_c}$ is larger than that of $\sigma_{\rm NNLO,\mu_\Lambda=1 GeV}$.}.

\section{Summary} \label{III}

\begin{figure}[htbp]
\centering
\includegraphics[width=0.8\textwidth]{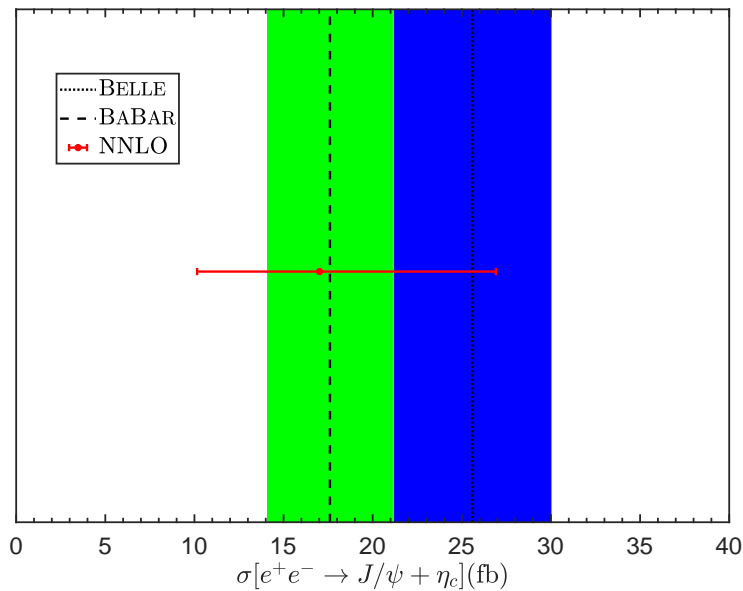}
\caption{The comparison between the NNLO QCD correction to $e^+e^- \to J/\psi+\eta_c$ and the experimental measurements} \label{errors}
\end{figure}

As a summary, we have calculated the NNLO QCD corrections of $J/\psi+\eta_c$ production in $e^+e^-$ annihilation at center-of-mass energy $\sqrt{s}=10.58$ GeV. The NNLO corrections to the total cross section for $e^+e^- \to J/\psi + \eta_c$ is sizable, but not comparable to the NLO corrections. It exhibits reasonable perturbative convergence behavior. More explicitly, the total cross section for $e^+e^- \to J/\psi + \eta_c$ is enhanced by $17\%$ in the case of $\mu_R=2m_c$ after including the NNLO corrections. From Fig.~\ref{result}, we find that the $\mu_R$ dependence of the total cross section is reduced at the NNLO level for $\mu_\Lambda=m_c$, and the predicted cross section is sensitive to the charm quark pole mass. Combining the results shown in Table~\ref{muLambdamc} and \ref{diffmc}, we obtain the final NNLO QCD corrections to $J/\psi+\eta_c$ production at the $B$ factories, i.e.,
\begin{eqnarray}
\sigma_{\rm NNLO}&=& 17.02^{+9.58+0+2.44}_{-6.14-3.09-0} \nonumber \\
&=& 17.02^{+9.89}_{-6.87} ~({\rm fb})
\end{eqnarray}
where the central value is obtained by taking $m_c=1.5 {\rm GeV}$, $\mu_R=2m_c$ and $\mu_\Lambda=m_c$. The first uncertainty is estimated by varying $m_c \in [1.3, 1.7] {\rm GeV}$, the second uncertainty is caused by varying $\mu_R \in [2m_c, \sqrt{s}/2]$ and the third uncertainty is obtained by varying $\mu_\Lambda \in [1 {\rm GeV}, m_c]$. Fig.~\ref{errors} shows that this prediction can both overlap with the B{\footnotesize A}B{\footnotesize AR} and B{\footnotesize ELLE} measurements within errors. The following work of this process is focus on improving the $\mu_R$ dependence.

\hspace{2cm}

\noindent {\bf Acknowledgments:} We would like to thank Yan-Qing Ma for sharing helpful unpublished code and Huai-Min Yu for helpful discussions. This work was supported by the National Natural Science Foundation of China with Grant Nos. 12135013, 11975242 and 12247129. It was also supported in part by National Key Research and Development Program of China under Contract No. 2020YFA0406400.

\hspace{2cm}

\bibliographystyle{JHEP}
\bibliography{nrqcd}

\providecommand{\href}[2]{#2}\begingroup\raggedright\begin{thebibliography}{10}

\bibitem{Fritzsch:1977ay}
H.~Fritzsch, {\it {Producing Heavy Quark Flavors in Hadronic Collisions: A Test
  of Quantum Chromodynamics}},  {\em Phys. Lett. B} {\bf 67} (1977) 217--221.

\bibitem{Halzen:1977rs}
F.~Halzen, {\it {Cvc for Gluons and Hadroproduction of Quark Flavors}},  {\em
  Phys. Lett. B} {\bf 69} (1977) 105--108.

\bibitem{Chang:1979nn}
C.-H. Chang, {\it {Hadronic Production of $J/\psi$ Associated With a Gluon}},
  {\em Nucl. Phys. B} {\bf 172} (1980) 425--434.

\bibitem{Berger:1980ni}
E.~L. Berger and D.~L. Jones, {\it {Inelastic Photoproduction of J/psi and
  Upsilon by Gluons}},  {\em Phys. Rev. D} {\bf 23} (1981) 1521--1530.

\bibitem{Matsui:1986dk}
T.~Matsui and H.~Satz, {\it {$J/\psi$ Suppression by Quark-Gluon Plasma
  Formation}},  {\em Phys. Lett. B} {\bf 178} (1986) 416--422.

\bibitem{Bodwin:1994jh}
G.~T. Bodwin, E.~Braaten, and G.~P. Lepage, {\it {Rigorous QCD analysis of
  inclusive annihilation and production of heavy quarkonium}},  {\em Phys. Rev.
  D} {\bf 51} (1995) 1125--1171,
  [\href{http://arxiv.org/abs/hep-ph/9407339}{{\tt hep-ph/9407339}}]. [Erratum:
  Phys.Rev.D 55, 5853 (1997)].

\bibitem{Brambilla:2010cs}
N.~Brambilla et~al., {\it {Heavy Quarkonium: Progress, Puzzles, and
  Opportunities}},  {\em Eur. Phys. J. C} {\bf 71} (2011) 1534,
  [\href{http://arxiv.org/abs/1010.5827}{{\tt arXiv:1010.5827}}].

\bibitem{Andronic:2015wma}
A.~Andronic et~al., {\it {Heavy-flavour and quarkonium production in the LHC
  era: from proton\textendash{}proton to heavy-ion collisions}},  {\em Eur.
  Phys. J. C} {\bf 76} (2016), no.~3 107,
  [\href{http://arxiv.org/abs/1506.03981}{{\tt arXiv:1506.03981}}].

\bibitem{Lansberg:2019adr}
J.-P. Lansberg, {\it {New Observables in Inclusive Production of Quarkonia}},
  {\em Phys. Rept.} {\bf 889} (2020) 1--106,
  [\href{http://arxiv.org/abs/1903.09185}{{\tt arXiv:1903.09185}}].

\bibitem{Chen:2021tmf}
A.-P. Chen, Y.-Q. Ma, and H.~Zhang, {\it {A Short Theoretical Review of
  Charmonium Production}},  {\em Adv. High Energy Phys.} {\bf 2022} (2022)
  7475923, [\href{http://arxiv.org/abs/2109.04028}{{\tt arXiv:2109.04028}}].

\bibitem{Belle:2002tfa}
{\bf Belle} Collaboration, K.~Abe et~al., {\it {Observation of double c anti-c
  production in e+ e- annihilation at s**(1/2) approximately 10.6-GeV}},  {\em
  Phys. Rev. Lett.} {\bf 89} (2002) 142001,
  [\href{http://arxiv.org/abs/hep-ex/0205104}{{\tt hep-ex/0205104}}].

\bibitem{Belle:2004abn}
{\bf Belle} Collaboration, K.~Abe et~al., {\it {Study of double charmonium
  production in e+ e- annihilation at s**(1/2) \textasciitilde{} 10.6-GeV}},
  {\em Phys. Rev. D} {\bf 70} (2004) 071102,
  [\href{http://arxiv.org/abs/hep-ex/0407009}{{\tt hep-ex/0407009}}].

\bibitem{BaBar:2005nic}
{\bf BaBar} Collaboration, B.~Aubert et~al., {\it {Measurement of double
  charmonium production in $e^+e^-$ annihilations at $\sqrt{s}=10.6$ GeV}},
  {\em Phys. Rev. D} {\bf 72} (2005) 031101,
  [\href{http://arxiv.org/abs/hep-ex/0506062}{{\tt hep-ex/0506062}}].

\bibitem{Braaten:2002fi}
E.~Braaten and J.~Lee, {\it {Exclusive Double Charmonium Production from $e^+
  e^-$ Annihilation into a Virtual Photon}},  {\em Phys. Rev. D} {\bf 67}
  (2003) 054007, [\href{http://arxiv.org/abs/hep-ph/0211085}{{\tt
  hep-ph/0211085}}]. [Erratum: Phys.Rev.D 72, 099901 (2005)].

\bibitem{Liu:2002wq}
K.-Y. Liu, Z.-G. He, and K.-T. Chao, {\it {Problems of double charm production
  in e+ e- annihilation at s**(1/2) = 10.6-GeV}},  {\em Phys. Lett. B} {\bf
  557} (2003) 45--54, [\href{http://arxiv.org/abs/hep-ph/0211181}{{\tt
  hep-ph/0211181}}].

\bibitem{Hagiwara:2003cw}
K.~Hagiwara, E.~Kou, and C.-F. Qiao, {\it {Exclusive $J/\psi$ productions at
  $e^{+} e^{-}$ colliders}},  {\em Phys. Lett. B} {\bf 570} (2003) 39--45,
  [\href{http://arxiv.org/abs/hep-ph/0305102}{{\tt hep-ph/0305102}}].

\bibitem{Bodwin:2006ke}
G.~T. Bodwin, D.~Kang, T.~Kim, J.~Lee, and C.~Yu, {\it {Relativistic
  Corrections to e+ e- ---\ensuremath{>} J/psi + eta(c) in a Potential Model}},
   {\em AIP Conf. Proc.} {\bf 892} (2007), no.~1 315--317,
  [\href{http://arxiv.org/abs/hep-ph/0611002}{{\tt hep-ph/0611002}}].

\bibitem{He:2007te}
Z.-G. He, Y.~Fan, and K.-T. Chao, {\it {Relativistic corrections to J/psi
  exclusive and inclusive double charm production at B factories}},  {\em Phys.
  Rev. D} {\bf 75} (2007) 074011,
  [\href{http://arxiv.org/abs/hep-ph/0702239}{{\tt hep-ph/0702239}}].

\bibitem{Bodwin:2007ga}
G.~T. Bodwin, J.~Lee, and C.~Yu, {\it {Resummation of Relativistic Corrections
  to e+ e- ---\ensuremath{>} J/psi + eta(c)}},  {\em Phys. Rev. D} {\bf 77}
  (2008) 094018, [\href{http://arxiv.org/abs/0710.0995}{{\tt
  arXiv:0710.0995}}].

\bibitem{Ma:2004qf}
J.~P. Ma and Z.~G. Si, {\it {Predictions for e+ e- ---\ensuremath{>} J/psi
  eta(c) with light-cone wave-functions}},  {\em Phys. Rev. D} {\bf 70} (2004)
  074007, [\href{http://arxiv.org/abs/hep-ph/0405111}{{\tt hep-ph/0405111}}].

\bibitem{Bondar:2004sv}
A.~E. Bondar and V.~L. Chernyak, {\it {Is the BELLE result for the cross
  section sigma(e+ e- ---\ensuremath{>} J / psi + eta(c)) a real difficulty for
  QCD?}},  {\em Phys. Lett. B} {\bf 612} (2005) 215--222,
  [\href{http://arxiv.org/abs/hep-ph/0412335}{{\tt hep-ph/0412335}}].

\bibitem{Bodwin:2006dm}
G.~T. Bodwin, D.~Kang, and J.~Lee, {\it {Reconciling the light-cone and NRQCD
  approaches to calculating e+ e- ---\ensuremath{>} J/psi + eta(c)}},  {\em
  Phys. Rev. D} {\bf 74} (2006) 114028,
  [\href{http://arxiv.org/abs/hep-ph/0603185}{{\tt hep-ph/0603185}}].

\bibitem{Braguta:2008tg}
V.~V. Braguta, {\it {Double charmonium production at B-factories within light
  cone formalism}},  {\em Phys. Rev. D} {\bf 79} (2009) 074018,
  [\href{http://arxiv.org/abs/0811.2640}{{\tt arXiv:0811.2640}}].

\bibitem{Zeng:2021hwt}
L.~Zeng, H.-B. Fu, D.-D. Hu, L.-L. Chen, W.~Cheng, and X.-G. Wu, {\it
  {Revisiting the production of $J/\psi+\eta_c$ via the $e^+e^-$ annihilation
  within the QCD light-cone sum rules}},  {\em Phys. Rev. D} {\bf 103} (2021),
  no.~5 056012, [\href{http://arxiv.org/abs/2102.01842}{{\tt
  arXiv:2102.01842}}].

\bibitem{Zhang:2005cha}
Y.-J. Zhang, Y.-j. Gao, and K.-T. Chao, {\it {Next-to-leading order QCD
  correction to e+ e- ---\ensuremath{>} J / psi + eta(c) at s**(1/2) =
  10.6-GeV}},  {\em Phys. Rev. Lett.} {\bf 96} (2006) 092001,
  [\href{http://arxiv.org/abs/hep-ph/0506076}{{\tt hep-ph/0506076}}].

\bibitem{Gong:2007db}
B.~Gong and J.-X. Wang, {\it {QCD corrections to $J/\psi$ plus $\eta_c$
  production in $e^{+} e^{-}$ annihilation at $S^{(1/2)}$ = 10.6-GeV}},  {\em
  Phys. Rev. D} {\bf 77} (2008) 054028,
  [\href{http://arxiv.org/abs/0712.4220}{{\tt arXiv:0712.4220}}].

\bibitem{Dong:2012xx}
H.-R. Dong, F.~Feng, and Y.~Jia, {\it {$O(\alpha_s v^2)$ correction to
  $e^+e^-\to J/\psi+\eta_c$ at $B$ factories}},  {\em Phys. Rev. D} {\bf 85}
  (2012) 114018, [\href{http://arxiv.org/abs/1204.4128}{{\tt
  arXiv:1204.4128}}].

\bibitem{Li:2013otv}
X.-H. Li and J.-X. Wang, {\it {$O(\alpha_{s}\upsilon^{2})$ correction to
  $J/\psi$ plus $\eta_c$ production in $e^{+}e^{-}$ annihilation at $\sqrt{s}
  =$ 10.6 GeV}},  {\em Chin. Phys. C} {\bf 38} (2014) 043101,
  [\href{http://arxiv.org/abs/1301.0376}{{\tt arXiv:1301.0376}}].

\bibitem{Sun:2018rgx}
Z.~Sun, X.-G. Wu, Y.~Ma, and S.~J. Brodsky, {\it {Exclusive production of
  $J/\psi+\eta_c$ at the $B$ factories Belle and Babar using the principle of
  maximum conformality}},  {\em Phys. Rev. D} {\bf 98} (2018), no.~9 094001,
  [\href{http://arxiv.org/abs/1807.04503}{{\tt arXiv:1807.04503}}].

\bibitem{Brodsky:2013vpa}
S.~J. Brodsky, M.~Mojaza, and X.-G. Wu, {\it {Systematic Scale-Setting to All
  Orders: The Principle of Maximum Conformality and Commensurate Scale
  Relations}},  {\em Phys. Rev. D} {\bf 89} (2014) 014027,
  [\href{http://arxiv.org/abs/1304.4631}{{\tt arXiv:1304.4631}}].

\bibitem{Shen:2017pdu}
J.-M. Shen, X.-G. Wu, B.-L. Du, and S.~J. Brodsky, {\it {Novel All-Orders
  Single-Scale Approach to QCD Renormalization Scale-Setting}},  {\em Phys.
  Rev. D} {\bf 95} (2017), no.~9 094006,
  [\href{http://arxiv.org/abs/1701.08245}{{\tt arXiv:1701.08245}}].

\bibitem{Wu:2019mky}
X.-G. Wu, J.-M. Shen, B.-L. Du, X.-D. Huang, S.-Q. Wang, and S.~J. Brodsky,
  {\it {The QCD renormalization group equation and the elimination of
  fixed-order scheme-and-scale ambiguities using the principle of maximum
  conformality}},  {\em Prog. Part. Nucl. Phys.} {\bf 108} (2019) 103706,
  [\href{http://arxiv.org/abs/1903.12177}{{\tt arXiv:1903.12177}}].

\bibitem{Huang:2021hzr}
X.-D. Huang, J.~Yan, H.-H. Ma, L.~Di~Giustino, J.-M. Shen, X.-G. Wu, and S.~J.
  Brodsky, {\it {Detailed Comparison of Renormalization Scale-Setting
  Procedures based on the Principle of Maximum Conformality}},
  \href{http://arxiv.org/abs/2109.12356}{{\tt arXiv:2109.12356}}.

\bibitem{Feng:2019zmt}
F.~Feng, Y.~Jia, Z.~Mo, W.-L. Sang, and J.-Y. Zhang, {\it
  {Next-to-next-to-leading-order QCD corrections to $e^+e^-\to J/\psi+\eta_c$
  at $B$ factories}},  \href{http://arxiv.org/abs/1901.08447}{{\tt
  arXiv:1901.08447}}.

\bibitem{Liu:2017jxz}
X.~Liu, Y.-Q. Ma, and C.-Y. Wang, {\it {A Systematic and Efficient Method to
  Compute Multi-loop Master Integrals}},  {\em Phys. Lett. B} {\bf 779} (2018)
  353--357, [\href{http://arxiv.org/abs/1711.09572}{{\tt arXiv:1711.09572}}].

\bibitem{Liu:2020kpc}
X.~Liu, Y.-Q. Ma, W.~Tao, and P.~Zhang, {\it {Calculation of Feynman loop
  integration and phase-space integration via auxiliary mass flow}},  {\em
  Chin. Phys. C} {\bf 45} (2021), no.~1 013115,
  [\href{http://arxiv.org/abs/2009.07987}{{\tt arXiv:2009.07987}}].

\bibitem{Liu:2021wks}
X.~Liu and Y.-Q. Ma, {\it {Multiloop corrections for collider processes using
  auxiliary mass flow}},  {\em Phys. Rev. D} {\bf 105} (2022), no.~5 L051503,
  [\href{http://arxiv.org/abs/2107.01864}{{\tt arXiv:2107.01864}}].

\bibitem{Liu:2022chg}
X.~Liu and Y.-Q. Ma, {\it {AMFlow: A Mathematica package for Feynman integrals
  computation via auxiliary mass flow}},  {\em Comput. Phys. Commun.} {\bf 283}
  (2023) 108565, [\href{http://arxiv.org/abs/2201.11669}{{\tt
  arXiv:2201.11669}}].

\bibitem{Czarnecki:1997vz}
A.~Czarnecki and K.~Melnikov, {\it {Two loop QCD corrections to the heavy quark
  pair production cross-section in e+ e- annihilation near the threshold}},
  {\em Phys. Rev. Lett.} {\bf 80} (1998) 2531--2534,
  [\href{http://arxiv.org/abs/hep-ph/9712222}{{\tt hep-ph/9712222}}].

\bibitem{Beneke:1997jm}
M.~Beneke, A.~Signer, and V.~A. Smirnov, {\it {Two loop correction to the
  leptonic decay of quarkonium}},  {\em Phys. Rev. Lett.} {\bf 80} (1998)
  2535--2538, [\href{http://arxiv.org/abs/hep-ph/9712302}{{\tt
  hep-ph/9712302}}].

\bibitem{Czarnecki:2001zc}
A.~Czarnecki and K.~Melnikov, {\it {Charmonium decays: J / psi
  ---\ensuremath{>} e+ e- and eta(c) ---\ensuremath{>} gamma gamma}},  {\em
  Phys. Lett. B} {\bf 519} (2001) 212--218,
  [\href{http://arxiv.org/abs/hep-ph/0109054}{{\tt hep-ph/0109054}}].

\bibitem{Kniehl:2006qw}
B.~A. Kniehl, A.~Onishchenko, J.~H. Piclum, and M.~Steinhauser, {\it {Two-loop
  matching coefficients for heavy quark currents}},  {\em Phys. Lett. B} {\bf
  638} (2006) 209--213, [\href{http://arxiv.org/abs/hep-ph/0604072}{{\tt
  hep-ph/0604072}}].

\bibitem{Hoang:2006ty}
A.~H. Hoang and P.~Ruiz-Femenia, {\it {Heavy pair production currents with
  general quantum numbers in dimensionally regularized NRQCD}},  {\em Phys.
  Rev. D} {\bf 74} (2006) 114016,
  [\href{http://arxiv.org/abs/hep-ph/0609151}{{\tt hep-ph/0609151}}].

\bibitem{Chung:2020zqc}
H.~S. Chung, {\it {$\overline {MS}$ renormalization of $S$-wave quarkonium
  wavefunctions at the origin}},  {\em JHEP} {\bf 12} (2020) 065,
  [\href{http://arxiv.org/abs/2007.01737}{{\tt arXiv:2007.01737}}].

\bibitem{CTEQ:1993hwr}
{\bf CTEQ} Collaboration, R.~Brock et~al., {\it {Handbook of perturbative QCD:
  Version 1.0}},  {\em Rev. Mod. Phys.} {\bf 67} (1995) 157--248.

\bibitem{Sun:2021tma}
Z.~Sun, {\it {Next-to-leading-order study of $J/\psi$ angular distributions in
  $e^{+}e^{-} \to J/\psi+\eta_c,\chi_{cJ}$ at $\sqrt{s} \approx 10.6$GeV}},
  {\em JHEP} {\bf 09} (2021) 073, [\href{http://arxiv.org/abs/2107.02047}{{\tt
  arXiv:2107.02047}}].

\bibitem{Gong:2009ng}
B.~Gong and J.-X. Wang, {\it {Next-to-leading-order QCD corrections to e+e-
  --\ensuremath{>} J/psi(cc) at the B factories}},  {\em Phys. Rev. D} {\bf 80}
  (2009) 054015, [\href{http://arxiv.org/abs/0904.1103}{{\tt
  arXiv:0904.1103}}].

\bibitem{Hahn:2000kx}
T.~Hahn, {\it {Generating Feynman diagrams and amplitudes with FeynArts 3}},
  {\em Comput. Phys. Commun.} {\bf 140} (2001) 418--431,
  [\href{http://arxiv.org/abs/hep-ph/0012260}{{\tt hep-ph/0012260}}].

\bibitem{Mertig:1990an}
R.~Mertig, M.~Bohm, and A.~Denner, {\it {FEYN CALC: Computer algebraic
  calculation of Feynman amplitudes}},  {\em Comput. Phys. Commun.} {\bf 64}
  (1991) 345--359.

\bibitem{Shtabovenko:2016sxi}
V.~Shtabovenko, R.~Mertig, and F.~Orellana, {\it {New Developments in FeynCalc
  9.0}},  {\em Comput. Phys. Commun.} {\bf 207} (2016) 432--444,
  [\href{http://arxiv.org/abs/1601.01167}{{\tt arXiv:1601.01167}}].

\bibitem{Klappert:2020nbg}
J.~Klappert, F.~Lange, P.~Maierh\"ofer, and J.~Usovitsch, {\it {Integral
  reduction with Kira 2.0 and finite field methods}},  {\em Comput. Phys.
  Commun.} {\bf 266} (2021) 108024,
  [\href{http://arxiv.org/abs/2008.06494}{{\tt arXiv:2008.06494}}].

\bibitem{Kramer:1995nb}
M.~Kr\"amer, {\it {QCD corrections to inelastic J / psi photoproduction}},
  {\em Nucl. Phys. B} {\bf 459} (1996) 3--50,
  [\href{http://arxiv.org/abs/hep-ph/9508409}{{\tt hep-ph/9508409}}].

\bibitem{Broadhurst:1991fy}
D.~J. Broadhurst, N.~Gray, and K.~Schilcher, {\it {Gauge invariant on-shell
  Z(2) in QED, QCD and the effective field theory of a static quark}},  {\em Z.
  Phys. C} {\bf 52} (1991) 111--122.

\bibitem{Bekavac:2007tk}
S.~Bekavac, A.~Grozin, D.~Seidel, and M.~Steinhauser, {\it {Light quark mass
  effects in the on-shell renormalization constants}},  {\em JHEP} {\bf 10}
  (2007) 006, [\href{http://arxiv.org/abs/0708.1729}{{\tt arXiv:0708.1729}}].

\bibitem{Czakon:2007ej}
M.~Czakon, A.~Mitov, and S.~Moch, {\it {Heavy-quark production in massless
  quark scattering at two loops in QCD}},  {\em Phys. Lett. B} {\bf 651} (2007)
  147--159, [\href{http://arxiv.org/abs/0705.1975}{{\tt arXiv:0705.1975}}].

\bibitem{Czakon:2007wk}
M.~Czakon, A.~Mitov, and S.~Moch, {\it {Heavy-quark production in gluon fusion
  at two loops in QCD}},  {\em Nucl. Phys. B} {\bf 798} (2008) 210--250,
  [\href{http://arxiv.org/abs/0707.4139}{{\tt arXiv:0707.4139}}].

\bibitem{Fael:2020bgs}
M.~Fael, K.~Sch\"onwald, and M.~Steinhauser, {\it {Exact results for $
  {Z}_m^{\mathrm{OS}} $ and $ {Z}_2^{\mathrm{OS}} $ with two mass scales and up
  to three loops}},  {\em JHEP} {\bf 10} (2020) 087,
  [\href{http://arxiv.org/abs/2008.01102}{{\tt arXiv:2008.01102}}].

\bibitem{Barnreuther:2013qvf}
P.~B\"arnreuther, M.~Czakon, and P.~Fiedler, {\it {Virtual amplitudes and
  threshold behaviour of hadronic top-quark pair-production cross sections}},
  {\em JHEP} {\bf 02} (2014) 078, [\href{http://arxiv.org/abs/1312.6279}{{\tt
  arXiv:1312.6279}}].

\bibitem{Tao:2022qxa}
W.~Tao, R.~Zhu, and Z.-J. Xiao, {\it {Next-to-next-to-leading order matching of
  beauty-charmed meson $B_{c}$ and $B^*_{c}$ decay constants}},
  \href{http://arxiv.org/abs/2209.15521}{{\tt arXiv:2209.15521}}.

\bibitem{ParticleDataGroup:2022pth}
{\bf Particle Data Group} Collaboration, R.~L. Workman et~al., {\it {Review of
  Particle Physics}},  {\em PTEP} {\bf 2022} (2022) 083C01.

\bibitem{Herren:2017osy}
F.~Herren and M.~Steinhauser, {\it {Version 3 of RunDec and CRunDec}},  {\em
  Comput. Phys. Commun.} {\bf 224} (2018) 333--345,
  [\href{http://arxiv.org/abs/1703.03751}{{\tt arXiv:1703.03751}}].

\bibitem{Yu:2020tri}
H.-M. Yu, W.-L. Sang, X.-D. Huang, J.~Zeng, X.-G. Wu, and S.~J. Brodsky, {\it
  {Scale-fixed predictions for $\gamma + \eta_c$ production in
  electron-positron collisions at NNLO in perturbative QCD}},  {\em JHEP} {\bf
  01} (2021) 131, [\href{http://arxiv.org/abs/2007.14553}{{\tt
  arXiv:2007.14553}}].

\end{thebibliography}\endgroup

\end{document}